\documentclass{article}

\usepackage{graphicx}
\usepackage{hyperref}
\usepackage{authblk}
\usepackage{float}
\usepackage{geometry}
\usepackage{multicol}
\usepackage{amsmath}
\usepackage{bm}% bold math
\usepackage{braket}
\usepackage[utf8]{inputenc}
\usepackage{cite}
\usepackage{authblk}
\usepackage{color}
\usepackage{soul}
\soulregister\cite{7} % make \hl{} compatible with \cite
\soulregister\ref{7} % make \hl{} compatible with \ref
\soulregister\ref{7} % make \hl{} compatible with \ref
\usepackage[framed,numbered,autolinebreaks,useliterate]{mcode}
\newcommand\nnfootnote[1]{%
  \begin{NoHyper}
  \renewcommand\thefootnote{}\footnote{$\dagger$ The first two authors contributed equally to this work.}%
  \addtocounter{footnote}{-1}%
  \end{NoHyper}
}
\begin{document} \sloppy

\title{Soliton-pair dynamical transition in mode-locked lasers}
\author[1,$\dagger$]{Kfir Sulimany}
\author[1,$\dagger$]{Offek Tziperman}
\author[1]{Yaron Bromberg}
\author[1,2]{Omri Gat}
\affil[1]{Racah Institute of Physics, The Hebrew University of Jerusalem, Jerusalem 91904, Israel}
\affil[2]{Corresponding author: omrigat@mail.huji.ac.il}

\date{}
\twocolumn[\begin{@twocolumnfalse}
\maketitle
\begin{abstract}
Multi-soliton mode-locked laser waveforms are much sought as a complex light source for research and applications, but are difficult to manipulate effectively because of the elaborate and diverse interactions present. Here we present an experimental, numerical, and theoretical study of the interaction and control of the internal dynamics of a two-soliton waveform in a mode-locked fiber laser. Using the pumping current as a control agent, we demonstrate experimentally a two-orders-of-magnitude reduction in the separation of a bound soliton pair, inducing a dynamical transition between a loosely bound, phase-incoherent pair, and a tightly bound phase-locked pair. We show on the basis of a Haus-model numerical simulation of the recently-proposed noise-mediated interaction theory, that the pulse separation and dynamical transition are governed by the shape of the dispersive-wave pedestals. We explain the dynamical transition by showing analytically, within a simplified theory, that the noise-mediated interaction becomes purely attractive when the pedestals energy drops below a threshold. This work demonstrates the ability to control the waveform through the interaction forces, without external intervention in the light propagation in the laser.
\end{abstract}

\end{@twocolumnfalse}]
\section{Introduction}

Dissipative solitons are localized waves in open systems far from equilibrium, whose existence results from a balance of dissipative, nonlinear, and dispersive effects. They appear in numerous physical areas including reaction-diffusion systems, neurological and ecological sciences, fluid dynamics, and photonics \cite{purwins2010dissipative, akhmediev2005dissipative, liehr2013dissipative, grelu2012dissipative, grelu2015nonlinear, song2019recent, javaloyes2016dynamics}. These nonlinear systems, which exhibit a plethora of unique nonlinear phenomena, are unified by the presence of solitons. Photonics provides an outstanding platform to study complex nonlinear interactions of dissipative solitons, which naturally emerge in strongly-pumped passively mode-locked lasers  \cite{ryczkowski2018real, liu2019revealing, liu2019revealing1, peng2019soliton, ning2014vector, wang2018self,marconi2014lasing}. In particular, collective nonlinear phenomena such as the aggregation of solitons into bunches have created great interest due to their potential applications in optical communication systems, and their resemblance to molecules \cite{richardson320FsSoliton1991,grudininPassiveHarmonicModelocking1993,pilipetskii1995acoustic,grudinin1997passive,kutz1998stabilized,tangCompoundPulseSolitons2003,tang2005soliton, tsatourian2013polarisation, wang2019optical, komarov2009quantization, zhao2009bunch, song2016coexistence, wang2018few, mou2013bound}. \nnfootnote

Soliton molecules and other multi-soliton patterns exhibit an extensive pallet of short- and long-range pulse interactions. Short-range interactions take place when pulse tails overlap \cite{malomed1995bound,akhmediev1998stable,grelu2002phase,lederer1999multipulse, tang2005soliton,olivier2009origin}. However, interactions over distances that are orders of magnitude longer than the width of individual solitons must be mediated by, for example, the gain medium \cite{kutz1998stabilized,weill2016noise} or by an acoustic response \cite{pilipetskii1995acoustic,jang2013ultraweak,he2019formation}. Furthermore, non-soliton components of optical pulses, such as dispersive-wave pedestals can give rise to interactions with an intermediate range, significantly longer than the soliton width, but much shorter than the cavity length \cite{grudinin1997passive,soto2003quantized}. Recently, optical modulation components have been utilized to induce transitions between complex multi-soliton patterns  \cite{he2019formation,zhou2020buildup,nimmesgern2021soliton}. However, a dynamical quasi-static transition between incoherently coupled solitons, separated by hundreds of times their width, and coherently coupled overlapping solitons, has not been reported to date. It is desirable to realize a laser with a straightforward and reproducible protocol for driving such transitions, supported by a theoretical underpinning.
%In addition, a model that captures the dynamics of both types of bound states is still lacking. A dynamical transition, accompanied by the proposed model, allow accurate control of coupled solitons over three orders of magnitude in separation, and a new qualitative understanding of the mechanism involved.

Recently, we introduced a long-range interaction mechanism arising from the effect of gain depletion in the presence of noisy quasi-CW light in mode-locked lasers \cite{weill2016noise, sulimany2018bidirectional}. The interaction results from the reduction of optical fluctuations due to gain depletion following the passage of a pulse through the gain medium. Suppression of fluctuations decreases the temporal jitter of subsequent pulses and in this way biases the jitter and sets a pulse drift motion. This noise-mediated interaction (NMI) mechanism shares some intriguing properties with the Casimir effect in quantum electrodynamics, where macroscopic objects experience an effective interaction due to the suppression of electromagnetic field fluctuations \cite{casimir1948influence}. In both the NMI and the Casimir effect, distant objects inhibit microscopic fluctuations in extended electromagnetic modes. The consequent breaking of spatiotemporal homogeneity gives rise to the weak interactions among these objects. 

Optical solitons in anomalous-dispersion mode-locked lasers are often accompanied by weak and broad dispersive-wave pedestals \cite{weill2011spectral}. When solitons are well-separated, the NMI is generally attractive \cite{weill2016noise}, but when two solitons are close enough that their pedestals overlap, the interaction can become repulsive; consequently, NMI can lead to the formation of bound states of solitons whose steady state separation is comparable with the pedestal width. The phases of solitons in such bound states are not coherently locked, so they can be viewed as \emph{loosely bound}. In contrast, when both the temporal separation and the relative phase of soliton pairs are locked, they are \emph{tightly bound} by coherent overlap interactions. In practice, the complex (1+1)-dimensional space-time nonlinear dynamics is driven by multiple interaction mechanisms where both time scales and space scales span many orders of magnitudes. Therefore, a precise control over the separation of bunched solitons is still lacking. 

\begin{figure}[t!]
\begin{centering}
\includegraphics[width=\columnwidth]{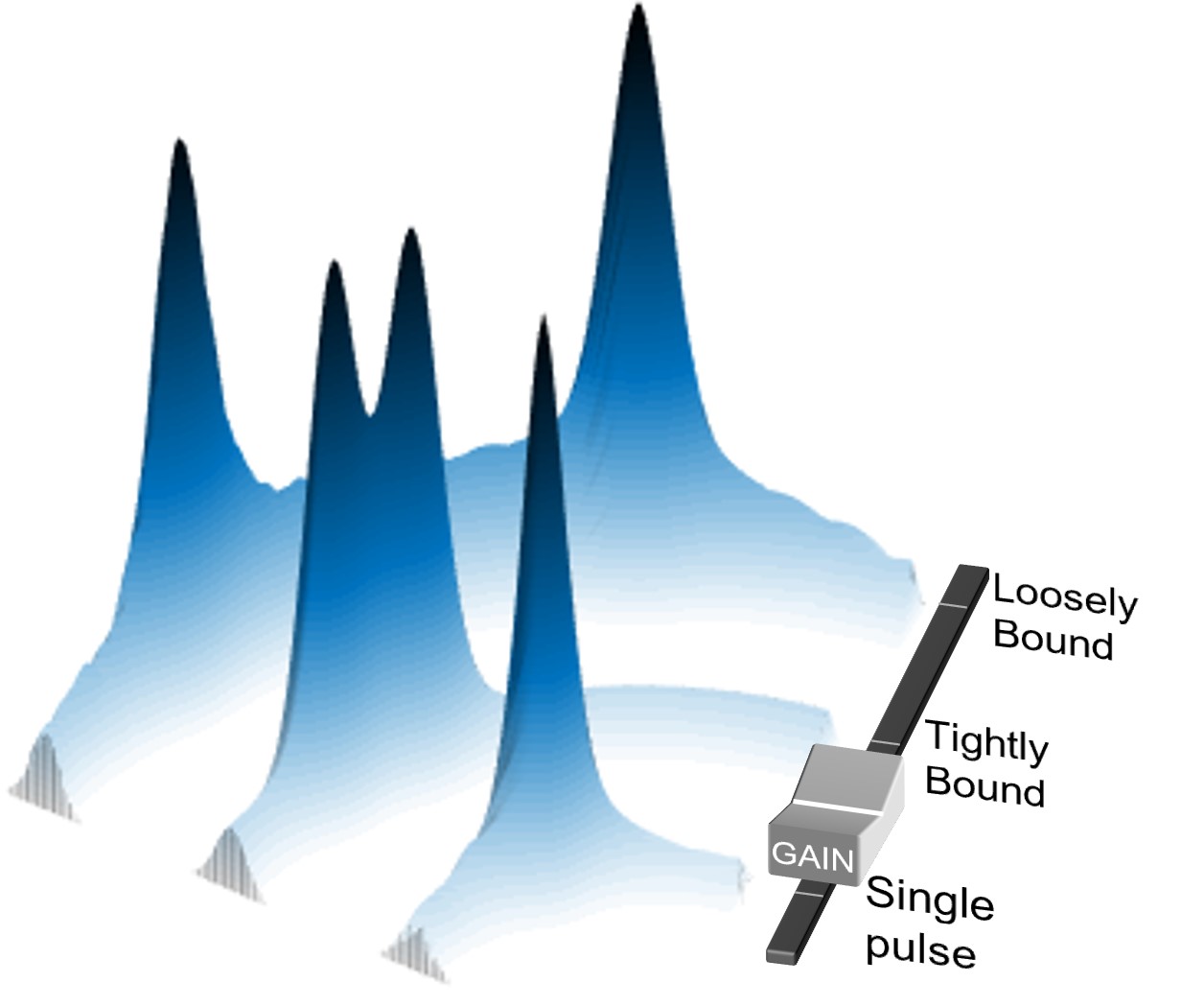}
\par\end{centering}
     \caption{\textbf{Soliton pair dynamical transition in a mode-locked laser}. The laser gain is used as a knob to control a dynamical transition between loosely and tightly bound soliton pairs (central and outer arcs). Above the gain threshold for stable single-pulse operation (inner arc), a steady state of two solitons is obtained. At high gain values, a loosely bound state (outer arc) is obtained as a result of the noise mediated interaction mechanism, which is driven by the pedestals around the solitons. By decreasing the gain, the soliton separation time decreases over a range spanning hundreds of times their duration and at an accuracy higher than the soliton duration. At lower gain values, the soliton separation becomes comparable to the duration of each soliton, and a dynamical transition to a tightly bound state (central arc) is observed.}
 \label{concept}

 \end{figure}

Here, we experimentally observe a dynamical transition between loosely- and tightly-bound soliton states in a passively mode-locked fiber laser. We use the pumping current as a control knob that drives the transition as illustrated in Figure \ref{concept}. When we pump the laser above the gain threshold for single-pulse operation (inner arc), a pair of solitons is formed \cite{bale2009transition} (central and outer arcs). At high gain values the solitons are loosely bound (outer arc) by the NMI, at a separation time dictated by their pedestals as shown in Figure \ref{concept}. We then lower the pumping current and observe a reduction of the solitons' separation time as a result of the reduction of the pedestals energy. We achieve a control of the separation time over a range of hundreds of times the solitons duration with an accuracy higher than the soliton duration. When the soliton separation drops under a threshold, the coherent interaction becomes dominant and drives the dynamical transition to a tightly bound state (central arc). 

We show that the transition is induced by NMI using a stochastic master equation that we developed for capturing the NMI mechanism, which reproduces the experimentally observed dynamical transition. Furthermore, we use a simplified analytical model of NMI to demonstrate that the transition point is set by the energy carried by the pedestals and by their temporal width. Since both parameters depend on the gain coefficient, which itself depends on the pumping current, it is possible to actively switch between loosely- and tightly-bound solitons by simply tuning the  
pumping current of the laser, providing a significant step towards applications of multipulse laser waveforms.

\section{Dynamical transition in a mode-locked fiber laser}

\subsection{Experimental observation}

\begin{figure}[t!]
\begin{centering}
\includegraphics[width=\columnwidth]{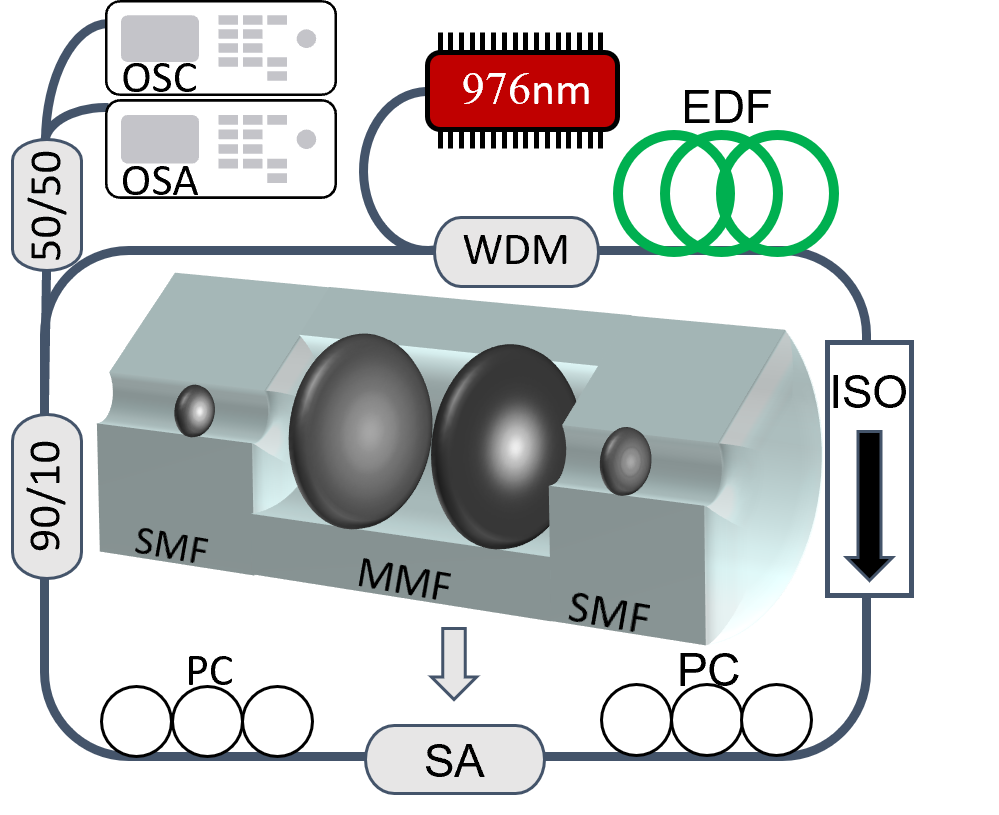}
\par\end{centering}
     \caption{\textbf{The mode-locked fiber laser used for studying the dynamical transition between loosely and tightly bound solitons.} The cavity length is 14\,m, including a 2.7\,m Erbium-doped fiber gain medium. The Erbium-doped fiber is core pumped by a 976\,nm diode laser through a wavelength-division multiplexer. Passive mode-locking is achieved by adding a 1.5\,m long segment of a graded-index multimode fiber. A polarization-independent optical isolator ensures unidirectional lasing, two fiber polarization controllers tune the overall low cavity birefringence and a 90/10 coupler provides the laser output. The laser output power is measured by a fast photodiode and analyzed with an optical spectrum analyzer and a real-time oscilloscope. SA: saturable absorber, SMF: single mode fiber, MMF: multimode fiber, EDF: Erbium-doped fiber, WDM: wavelength-division multiplexer, ISO: optical isolator, PC: polarization controller, OSA: optical spectrum analyzer, OSC: oscilloscope.}
 \label{exp_setup}
 \end{figure}

To experimentally study the dynamics of soliton pairs in laser cavities we constructed a mode-locked laser that is based on an all-fiber integrated ring cavity operated in the anomalous dispersion regime \cite{zhao2020observation}. Passive mode-locking is achieved by adding a segment of a multimode fiber (MMF) to the single-mode fiber cavity, as depicted in Figure \ref{exp_setup} \cite{fu2015passive, li2017mode, wang2017er, wang2018stretched, teugin2018all, yang2018saturable, chen2018all, zhao2020generation}. Since light entering the MMF can excite multiple fiber modes, the fraction of the light that is coupled back to the single mode fiber depends on the interference of the excited fiber modes. At high intensities the interference is modified by Kerr nonlinearity, making the transmission through the MMF intensity-dependent. By tweaking the properties and conformation of the MMF, the intensity-dependent transmission can be adjusted to realize a saturable absorber, with a controllable amount of linear loss, saturated loss, and saturation power. These additional degrees of freedom in the saturable absorber enable us to realize a wide range of multi-soliton waveforms \cite{zhao2018experimental}. For more details regarding the experimental methods and components used in this work please see supplementary information.
 
\begin{figure}[ht]
\begin{centering}
\includegraphics[width=\columnwidth]{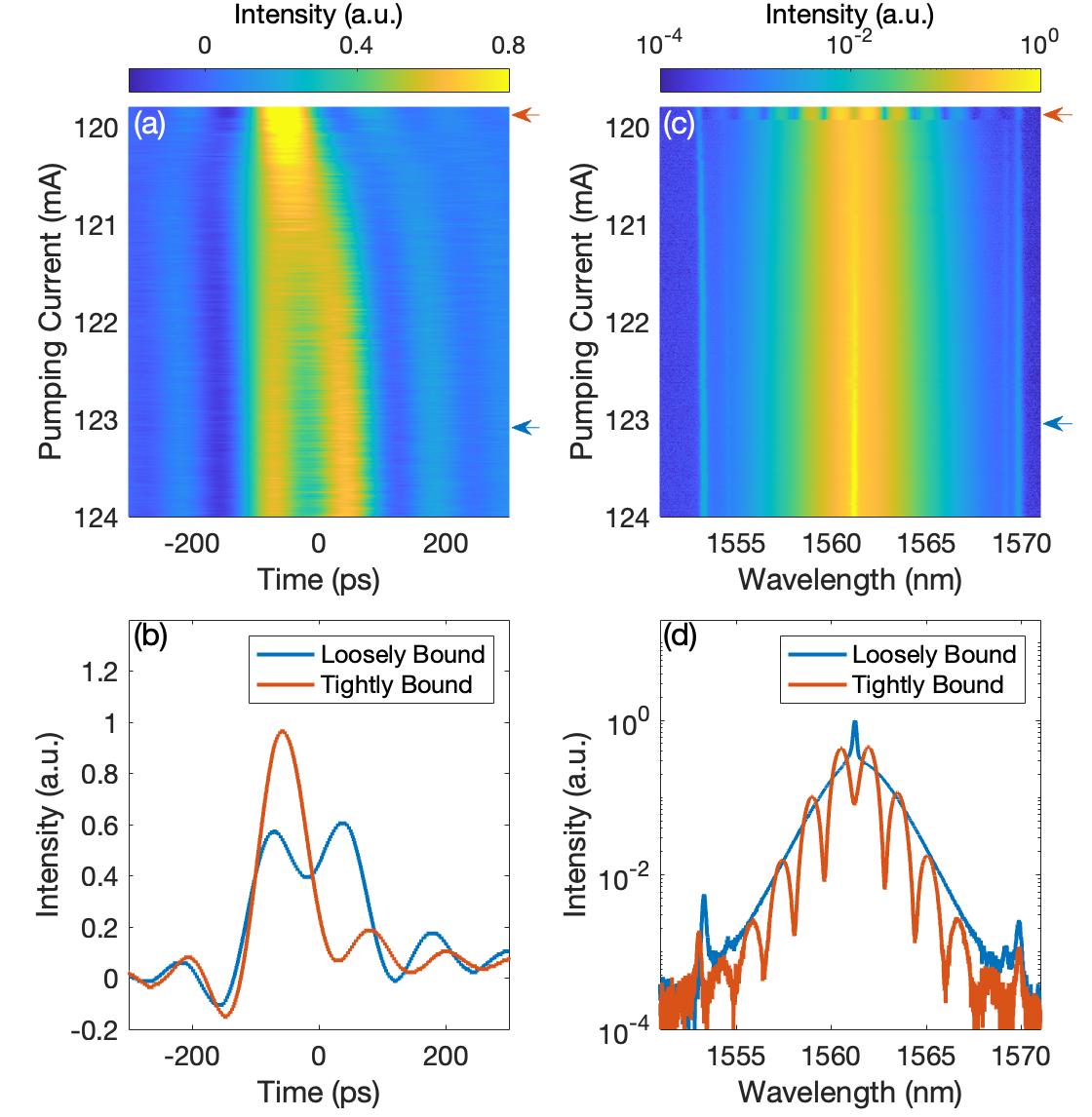}
\par\end{centering}
     \caption{\textbf{Soliton dynamical transition in a mode-locked fiber laser.} (a) The laser output temporal intensity profile measured as a function of the pumping current using a fast photodetector. Cross-sections at pumping currents of 123\,mA (blue arrows) and 120\,mA (red arrows), plotted in panel (b) with corresponding colors, demonstrate that the total energy in the soliton waveform remains nearly constant even when the inter-soliton separation is too small to be resolved by the oscilloscope. (c) The laser output spectrum measured as a function of the pumping current. Cross-sections at pumping currents of 123\,mA (blue arrows) and 120\,mA (red arrows), plotted in panel (d) with corresponding colors, demonstrate the two soliton binding regimes: loosely bound pairs with fluctuating relative phase and tightly bound pairs with coherently locked phases.}
  \label{transition_Exp}

 \end{figure}
 
Our experimental protocol starts by generating stationary states of soliton pairs by increasing the pumping current to 124\,mA, beyond the threshold for stable single-pulse operation. We then decrease the pumping current and study the dynamics of the soliton pairs. We start by studying the time-domain traces of the output intensity (Figure \ref{transition_Exp}(a)), exhibiting two peaks that approach each-other as the pumping current is decreased.  The soliton duration is sub-picosecond, estimated by the spectral profile presented below. However, the width of the pulses in Figure \ref{transition_Exp} (a) is determined by the temporal resolution of our detection apparatus, which is limited by the 8\,GHz bandwidth of the oscilloscope. For pumping currents below 121\,mA the soliton separation is too small to be resolved in Figure \ref{transition_Exp} (a), yet the approximate conservation of total energy in the soliton waveform indicates that the number of solitons has not changed, as expected from the clamping of the soliton energy in anomalous-dispersion passively mode-locked lasers \cite{grudininEnergyQuantisationFigure1992,Tang2005Mechanism,weillStatisticalLightmodeDynamics2007}. Figure \ref{transition_Exp} (b) shows examples of temporally resolved (blue) and unresolved (red) two-pulse waveform.

We continue by studying the output spectrum as a function of the pumping current (Figure \ref{transition_Exp}(c)). The central wavelength is 1561\,nm, and the full width at half maximum (FWHM) bandwidth is $\sim$3.8\,nm. At pumping currents above 120\,mA we do not observe spectral interference fringes, indicating that the solitons are not phase locked and they form a loosely bound state. At pumping currents below 120\,mA, we observe an abrupt appearance of spectral fringes, indicating that the solitons interact coherently so that their phases lock, i.e.\ the solitons are tightly bound. In both regimes, narrow sidebands appear at the primary Gordon-Kelly resonances \cite{kelly1992characteristic,gordonDispersivePerturbationsSolitons1992}, 1553\,nm and 1567\,nm, and their power grows with increasing pumping current. Figure \ref{transition_Exp}(d) presents two spectral cross-sections of Figure \ref{transition_Exp}(c), at pumping currents of 123\,mA (blue) and 120\,mA (red), exhibiting the Kelly sidebands and the interference fringes in the latter. Below 120\,mA in particular, the spacing of the spectral fringes implies that waveform consists of two solitons, with a separation time of a few picoseconds.

\subsection{Numerical simulations}
The temporal dynamics of dissipative solitons in fiber lasers is conveniently modelled by the Haus master equation for mode-locked lasers \cite{haus2000mode}. To show that the dynamical transition is caused by the NMI mechanism, we derived a generalized master equation that includes all of the physical processes needed to observe both the NMI and the coherent nonlinear interaction: 
  \begin{equation} \label{Haus}
 \begin{gathered} 
\frac{\partial u}{\partial z} = -\frac{l}{2}u-\frac{i\beta''}{2}\frac{\partial^2 u}{\partial t^2} + \frac{(g(u)+\kappa (z,t))}{2}(u+\frac{1}{\Omega _g^2}\frac{\partial^2u}{\partial t^2}) \\ 
+(i\gamma+\delta-\sigma |u|^2)|u|^2u + \eta (z,t)\ ,
\end{gathered}
\end{equation}
where $u=u(z,t)$ is the complex scalar envelope of the electric field, $t$ is the retarded time, $z$ is the propagation distance, $l$ is the linear loss coefficient, $\beta''$ is the group velocity dispersion, $g, \kappa$ are the gain coefficients to be described below, $\gamma$ is the Kerr coefficient, $\Omega_g$ is the gain bandwidth, and $\delta,\,\sigma$ are the fast saturable absorption constants, modeling bleaching and saturation of the absorber, respectively. $\eta (z,t)$ is a complex Gaussian white uncorrelated noise term, $\langle\eta(z,t)^*\eta(z',t')\rangle=\zeta\delta(z-z')\delta(t-t')$, where $\zeta$ is the noise strength. The noise term generates a non-homogeneous quasi-CW noise floor in the waveform that is the medium of the noise-mediated interaction mechanism. 
\begin{figure}[ht!]
\begin{centering}
\includegraphics[width=\columnwidth]{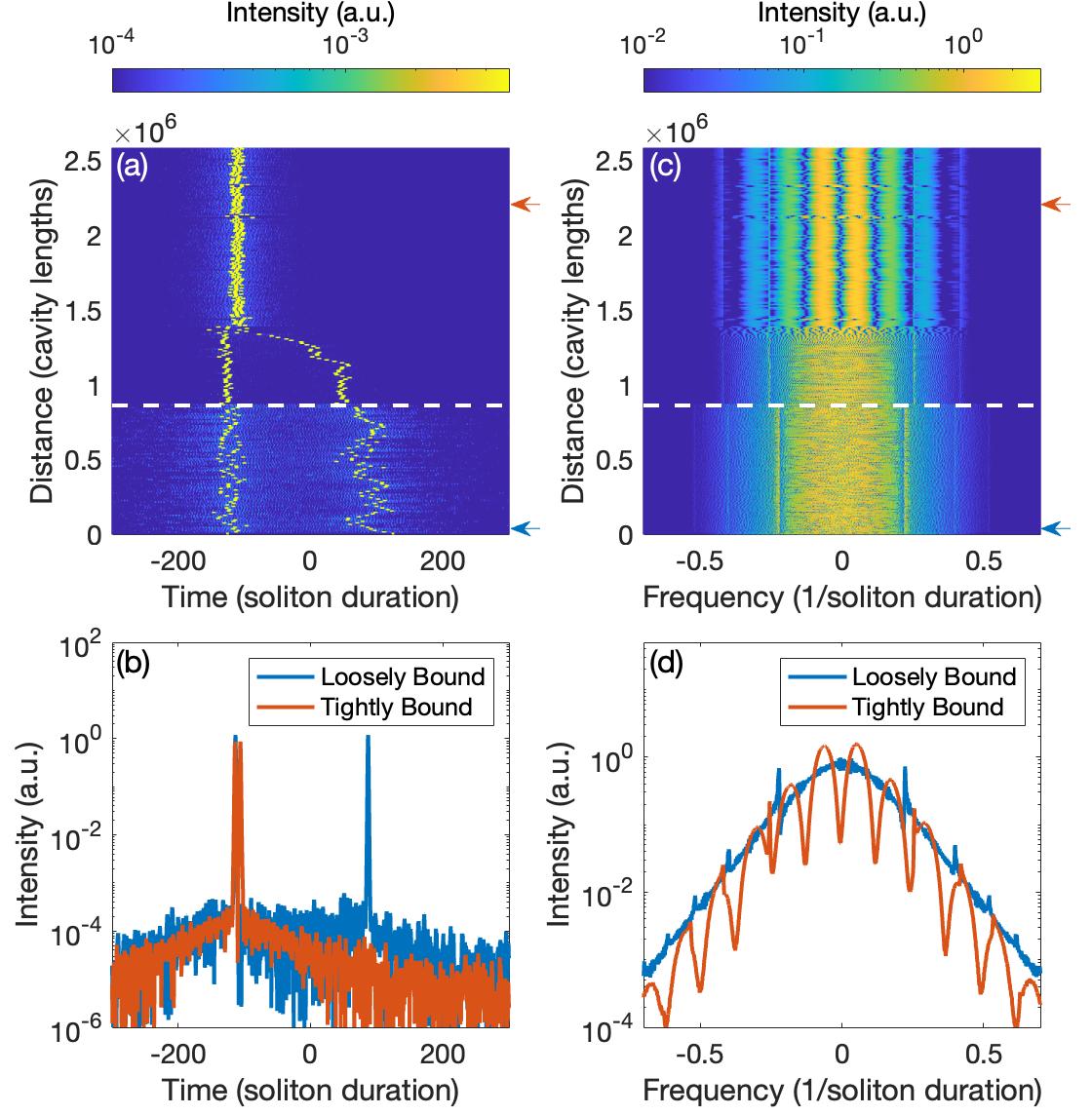}
\par\end{centering}
     \caption{\textbf{Soliton dynamical transition simulated by a stochastic Haus master equation.} (a) The simulated laser temporal intensity waveform as a function of propagation distance. After $8.7\cdot10^5$ roundtrips we reduced abruptly the unsaturated gain from $g_0=0.65$ to $g_0=0.58$; the high and low gain regimes are separated by the white dashed lines. Temporal cross-sections from before the unsaturated gain was changed (blue arrow) and after the tightly bound pair was formed (red arrow), are plotted in panel (b) with corresponding colors. The time traces exhibit around each soliton a symmetric, exponentially decaying pedestal, $\sim100\,$ times wider than the solitons. (c) The simulated laser output spectrum as a function of propagation distance. Cross-sections of the spectrum before the gain was changed (blue arrow) and after the solitons reached a tightly bound state (red arrow) are plotted in panel (d) with corresponding colors, demonstrating the two binding regimes: loosely bound pairs with fluctuating relative phase and tightly bound pairs with coherently locked phases. To depict the phase-locking each spectrum was averaged over 86 sequential traces spanning $8.6\cdot10^5$ cavity lengths.}
 \label{transition_sim}
 \end{figure}

Since the amplifier gain is depleted by the passage of a pulse, and gain recovery is slow, the depleted gain profile $g(u)+\kappa(z,t)$ is nonuniform, which means that the quasi-CW intensity is not uniform either. The depleted gain consists of two terms; the temporally-homogenous mean saturated gain $g(u)$ and the inhomogenous gain $\kappa$ that has zero mean and is a key element of the NMI interaction. If the relaxation of the gain medium is much slower than the pulse repetition rate, the mean saturated gain $g(u)$ can be well approximated by:
\begin{equation} \label{saturation}
g(u) = \frac{g_0}{1+P_{\text{av}} (u)/P_{\text{sat}}}
\end{equation}
where $g_0$ is the unsaturated gain, $P_{\text{sat}}$ is the saturation power of the amplifier, and $P_{\text{av}}(u)$ is the mean intracavity power
\begin{equation} \label{avarege}
P_\text{av} (u) = \frac{1}{t_R}  \int_{-t_R/2}^{t_R/2} |u(z,t)|^2dt 
\end{equation}
in which $t_R$ is the round-trip time.
The inhomogenous gain is governed by the equation
\begin{equation} \label{fluctuating}
\kappa (z,t) = -g_d \int_{-t_R/2}^{t} |u(z,t)|^2 dt + (t+\frac{t_R}{2})g_d P_\text{av}(u)  
\end{equation}
where $g_d$ is the gain depletion coefficient. To ensure that $\kappa(z,t)$ has a temporal mean of zero, we subtract its mean after using \eqref{fluctuating}. In Eq.\ \eqref{fluctuating} we approximate the recovery rate by a constant because the recovery time of the Erbium-doped fiber gain medium is much longer than the round-trip time \cite{hemanahausModelockingLasers2000}.

In the Haus model, the effects of the laser elements (nonlinear propagation, dispersion, gain, and saturable absorption) are averaged over one round trip in the laser cavity to give a continuous-variable equation that allows for faster simulation. However, it cannot reproduce the Gordon-Kelly sidebands \cite{kelly1992characteristic,gordonDispersivePerturbationsSolitons1992} that are essential for the dynamical transition. Thus, although we are not attempting to make a one-to-one numerical model of the experiment, we introduce an additional discrete loss component in each round trip, that models the cavity inhomogeneities that produce the dispersive-wave pedestals \cite{weill2011spectral}.

We use the split-step method to numerically integrate Eq. \eqref{Haus}; the numerical method is based on \cite{wang2013comparison, li2010geometrical} with the additional elements that drive the noise-mediated interaction as described below. The simulation parameters and a sketch of the code are given in the supplementary information, in addition to the complete open source Matlab package used in this work, which is available online \cite{soliton2022}. 

To observe the soliton-pair dynamical transition we apply the following protocol. First, we initiate the cavity with two solitons at an estimated steady state separation time. This estimation is based on simulations with different initial separations and unsaturated gain values. The unsaturated gain values are chosen well above the threshold for a stable soliton-pair state and slightly below the value for a formation of a third soliton. Next, we propagate the input field many cavity lengths, and then abruptly decrease the unsaturated gain to a value slightly above the value for which one of the solitons is annihilated. Such relatively low gain values support coherently locked states of soliton-pairs.

Figure \ref{transition_sim}(a) presents time-domain traces of the cavity waveform intensity as a function of the propagation distance. At the beginning of the propagation $g_0=0.65$, and the distance between the solitons jitters around their initial separation, approximately 200 times larger than their width, but there is no significant drift. The phases of the two solitons are not locked (see below), implying that the calculated waveform is in a loosely bound steady state. After $8.7\cdot10^5$ cavity lengths, we decrease abruptly the unsaturated gain to $g_0=0.58$, and the solitons start to drift towards each other, until a new, tightly bound, steady state is reached at $1.4\cdot10^6$ cavity lengths. Note that the simulated waveforms are snapshots of a trajectory between two steady states at two values of the unsaturated gain, in contrast to the experimental waveforms that are all in steady states

Decreasing the unsaturated gain also changes the amount of energy carried by the pedestals that surround each soliton. The pedestal energy drops by 75\% after the abrupt reduction in $g_0$ (Figure \ref{transition_sim}(b)). This drop is consistent with the experimentally observed changes in the Gordon-Kelly sidebands as the pumping current was reduced (see below), although the pedestal intensity was too weak to measure in our experiments.

Figure \ref{transition_sim}(c) presents the simulated output spectrum as a function of the propagation distance. The FWHM bandwidth of the soliton spectrum is $\sim$0.4. Narrow sidebands appear at the Gordon-Kelly resonances, of which the primary ones are detuned by $\pm0.2$ from the pulse peak. Their frequency detuning increases and their peak power decreases the moment we lower the unsaturated gain, as was also observed experimentally.
In contrast with the experimental data, where the spectrum is averaged over a few milliseconds, the numerical spectral traces are instantaneous, and therefore all of them exhibit  fringes of interference between the two soliton spectra. Therefore, to distinguish between phase-locked and unlocked states we compare averages of successive spectral traces in Figure \ref{transition_sim}(d). For the low unsaturated gain value we observe stable spectral fringes in the steady state, indicating that the solitons are tightly bound (red curve). In contrast, for the high unsaturated gain value we do not observe stable interference fringes (blue curve), so the pulse phases are not coherently locked; since the mean separation in this state is constant, the solitons are loosely bound. 

Our simulation paves the way for investigating many complex phenomena where the noise mediated interaction takes a major role. Here, we focus on the dynamical transition between the loosely and tightly bound states. We found this task to be much harder than simulating scenarios where only one of the interactions plays a significant role, such as when observing either a tightly or a loosely bound state, because of the large range of both the fast time scale and the slow time (space) scale, associated with the two types of bound states.

\subsection{Analytic model}
\begin{figure}[t!]
\begin{centering}

\includegraphics[width=\columnwidth]{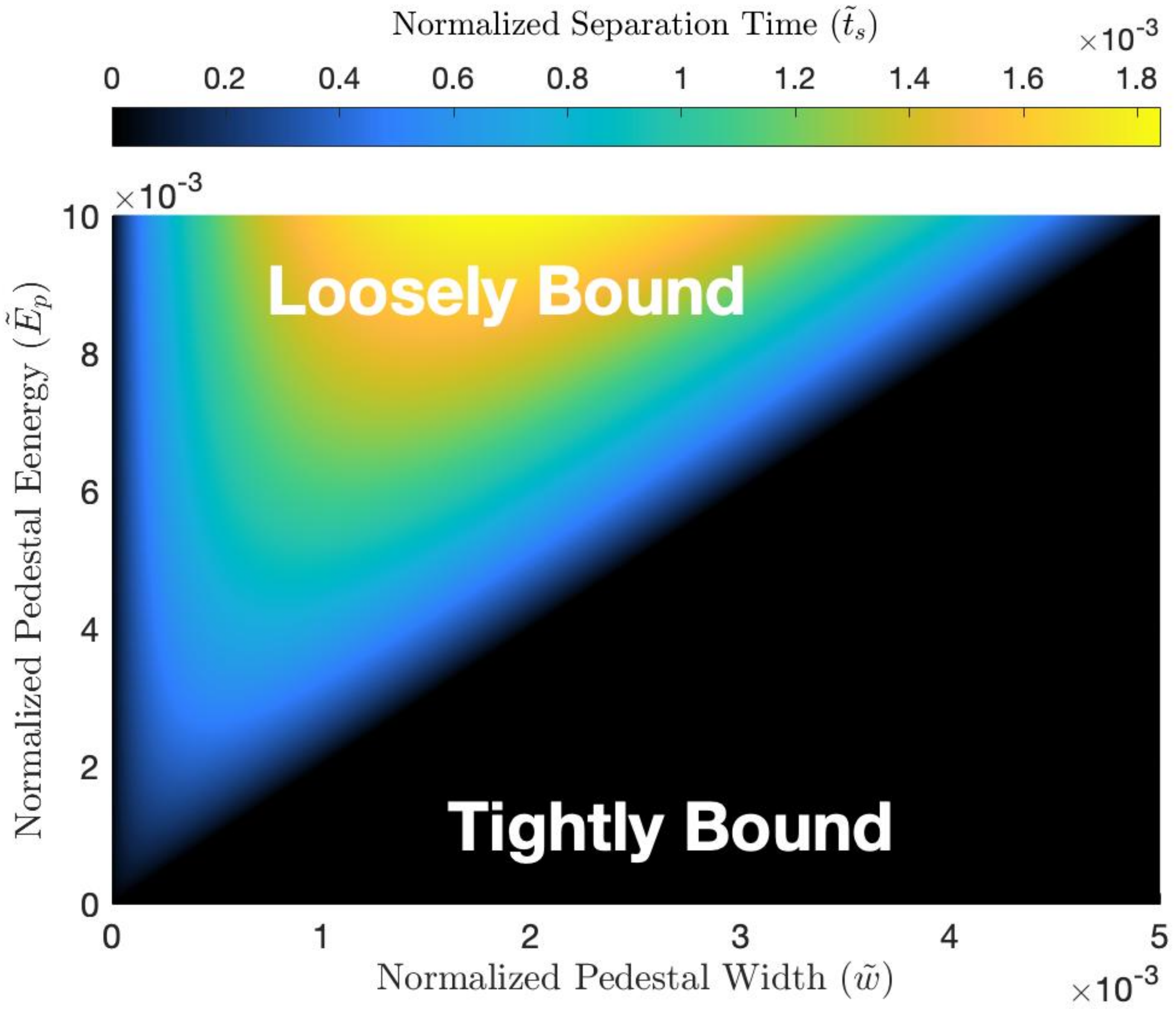}

\par\end{centering}
     \caption{ \textbf{Soliton steady states analysis.} 
      Theoretically predicted normalized separation time $\tilde t_s$ of the steady state of two solitons, normalized by the round trip time, plotted as a function of the pedestal width and energy, normalized by the total pulse energy, and the cavity round trip time respectively. Non-zero  values of $\tilde t_s$ are associated with loosely bound states, while $\tilde t_s=0$ corresponds to the regime where the NMI is always attractive and hence tightly bound steady states are observed (black region). The boundary between the $\tilde t_s>0$ and $\tilde t_s=0$ marks a dynamical transition line between loosely and tightly bound states. }
 \label{phase_diagram}
 \end{figure}

In order to obtain a simple expression for the loosely bound steady state separation time, we approximate the laser waveform by an incoherent sum of two non-overlapping soliton waveforms, each of which is accompanied by a weak and broad dispersive-wave pedestal generated by the Gordon-Kelly resonances \cite{weill2011spectral}. The soliton intensity profile is approximated by a delta function because its duration is much shorter than all other time scales, and the pedestal waveform is modeled by an exponential envelope decaying over a time scale $w$ that is much longer than the pulse width but much shorter than the cavity round-trip time $t_R$. The waveform intensity profile can therefore be approximated by
\begin{align} \label{pulses}
I(t)&=I_p(t-t_1(z))+I_p(t-t_2(z))\ ,\\
I_p(t)&=E_s\delta(t)+\frac{E_p}{2w}e^{-|t|/w}\ ,
\end{align}
$E_s$, $E_p$ being the pulse and pedestal energy, respectively. The NMI makes $t_1$ and $t_2$ drift under an effective potential until the mean separation $t_s=|\langle t_2-t_1 \rangle |$ reaches a steady state value that minimizes the effective potential \cite{weill2016noise} (see supplementary information for more details):

\begin{equation} \label{separation}
\tilde{t}_s=\tilde{w}\ln\Bigl( \frac{\tilde E_{p}}{2\tilde w}\Bigr)\ , \qquad \textnormal{if } \tilde E_{p}>2 \tilde{w}
\end{equation}
where $\tilde t_s=t_s/t_R$, $\tilde w=w/t_R$, and $\tilde E_p=E_p/(E_s+E_p)$. In contrast, if $\tilde{E}_{p}<2\tilde{w}$, the noise-mediated interaction is always attractive, so that the minimum of the effective potential appears at zero separation time $\tilde{t}_s=0$. Notice that Eq.\ \eqref{separation} corrects an algebraic mistake in the separation time formula that originally appeared in \cite{weill2016noise}.
Figure \ref{phase_diagram} is a color map showing the normalized steady state separation time $\tilde t_s$ as a function of the normalized pedestal width $\tilde w$ and the normalized pedestal energy $\tilde E_p$, for the typically small values of these variables. The black area represents zero steady state separation time, where the NMI is always attractive.

We next use these results to explain the dynamical transition between loosely and tightly bound soliton steady states observed in the experiments when the pumping current is lowered: when we  reduce the pumping current, the normalized pedestal energy decreases, which in turn shortens the separation time of the soliton pair, according to Eq.\ \eqref{separation}.
As the pumping current is lowered, therefore, two scenarios may occur. The first scenario is that the separation time decreases until the coherent interaction takes over at short separations. In this scenario, a dynamical transition occurs, after which the separation between the pulses, together with their relative phase, are locked by the coherent overlap interaction, as explained in the previous section. In the second scenario, when the pumping current is lowered, the soliton-pair state becomes unstable to pulse annihilation before the threshold for the transition between the loosely and tightly bound states is reached. 

\section{Discussion}
The principles of NMI are well in agreement with the experimental observations. Specifically, decreasing the pump power results in weaker pedestal energies and hence shorter separations of loosely bound solitons, where below some pumping threshold tightly bound pairs are formed. This hypothesis is strongly supported by the numerical simulation results, that are able to reproduce both the NMI drift of solitons toward loosely bound configurations for large gain values and the sharp diminishing of pedestal intensity profile, as well as the ensuing NMI attraction leading to the onset of overlap soliton interaction and the formation of tightly bound states when the gain is lowered.

The NMI mechanism is probably the main source of long-range inter-pulse interactions in most multipulse mode-locked fiber lasers. Thus, we seek to explain the dynamical transition between loosely and tightly bound soliton states in the NMI framework. In most complex dissipative systems, a theoretical modeling cannot provide accurate prediction for the transition parameters, since it often oversimplifies the complex dynamics in the system. In the present work, the NMI theory makes quantitative predictions for the soliton separation times in terms of the pedestal properties, but direct quantitative comparison of theory and experiment is not possible. One issue is that the existing NMI calculations are based on averaged pedestal width and energy, which in practice can oscillate significantly during a laser round trip, making the measurement of the average pedestal energy at the output coupler unreliable. More importantly, in our setup we could only estimate the pedestal duration indirectly from the width of Gordon-Kelly sidebands, obtaining a poor accuracy. Alternatively, the pedestal waveforms can be estimated for different pumping currents using the model presented in \cite{weill2011spectral}. Unfortunately, the prediction of this mode disagrees with our experimental and numerical simulations: While the theoretical model predicts that with weaker pumping the pedestal becomes temporally narrower, so that its energy decreases while its peak power remains fixed, in experiments and simulations we find that when the pumping current is decreased, the pedestal duration hardly changes, while the peak power decreases by a large factor. We attribute this discrepancy to the assumption of fixed soliton amplitude in \cite{weill2011spectral}. While it is well established that the amplitudes of solitons in multipulse mode-locked lasers are clamped to an optimal operation point \cite{weill2007statistical}, residual variations around it may remain, yielding large variations in the pedestal peak power.

\section{Conclusions and outlook}

This paper presents a dynamical transition of two-soliton waveforms in a mode-locked laser, in which the pulse separation time is governed by the long-range noise-mediated interaction (NMI) mechanism and a coherent overlap interaction. In the loosely bound pair regime, where the separation between the pulses ranges from a few tens to a few hundreds of picoseconds, the overlap interaction is negligible so that the soliton phases are not coherently locked, and the steady state separation is determined by the balance between long-range attraction and repulsion due to the dispersive-wave pedestals of the solitons.

When the pumping current is lowered, the pedestals become weaker, and the steady state separation between the solitons decreases; when the separation drops below a few dozen picoseconds, the coherent overlap interaction becomes effective and dominates over the NMI, binding the solitons tightly, so that their relative phases are locked and their separation is fixed. By tweaking the properties of the saturable absorber, one can control whether a dynamical transition between  loosely and tightly bound states can be observed or not: if the pumping current required for forming tightly bound states is too low for supporting two solitons in the cavity, the transition will not be observed. In this way, the NMI provides a systematic way to control the separation between the pulses in a two-pulse mode-locked laser configuration, and in particular, the transition between loosely and tightly bound configurations.

The theoretical explanation of the dynamical transition is based on numerical results and on analytic results derived from the NMI model of \cite{weill2016noise}. The full nonlinear dynamics are complex, and neither the numerical nor the analytic model capture all its details. Nevertheless, the key effects that we describe, namely the long-range NMI attraction, the pedestal-induced repulsion, the reduction of pedestal energy for smaller gain and consequent destabilization of the loosely bound configurations are independent of the fine details of the experiment, and have all been confirmed numerically. We used the simplified NMI model to show that the steady state soliton pair separation decreases to zero when the pedestal energy decreases to a finite value, confirming that unless a pulse is annihilated, reduction of pumping current eventually leads to the dynamical transition from a loosely to a tightly bound configuration.

Our experimental and numerical results show that the current theoretical understanding of dispersive wave dynamics in laser cavities, and the NMI effective interactions, are incomplete. We believe that the theoretical models can be improved by taking into account effects that have been neglected so far, such as the variations of the pulse amplitude as a function of the gain, oscillations of pedestal waveforms, and the finite response time of the quasi-CW to the gain depletion, and in this way achieve quantitative agreement between theory, numerical simulations and experiments. This will enable precise control of the soliton separation and the dynamical transition with the pump current. Furthermore, our numerical model provides a generic tool for simulating the dynamics of multipulse configurations in mode-locked lasers under different mechanisms of interaction.

Experimentally, better control of multi-soliton waveforms can be achieved by directly modifying the pedestal waveform with a reconfigurable spectral filter introduced into the laser cavity. Recently we have developed such a filter by engineering the transmission matrix of a multimode fiber, in an all-fiber configuration \cite{finkelstein2021spectral}. Such technology may allow fine control of the pulse separations without modifying the laser gain. The dynamical transition, along with the new understanding presented here of the interplay between the long and short range interactions will allow new novel ways to control multipulse bunches, providing a significant step towards the goal of multipulse waveform engineering.

\section{Acknowledgments}
The authors kindly thank Ori Katz for fruitful discussions and suggestions. 
This research was supported by the \textit{United States-Israel Binational Science Foundation (BSF)} (Grant No. 2017694) and the \textit{Israel Science Foundation (ISF)} (Grant No. 2403/20). KS and YB acknowledge the support of the Israeli Council for Higher Education, and the Zuckerman STEM Leadership Program. 

\bibliography{main.bib}

\begin{thebibliography}{10}

\bibitem{purwins2010dissipative}
H-G Purwins, HU~B{\"o}deker, and Sh~Amiranashvili.
\newblock Dissipative solitons.
\newblock {\em Advances in Physics}, 59(5):485--701, 2010.

\bibitem{akhmediev2005dissipative}
Neil Akhmediev and Adrian Ankiewicz.
\newblock Dissipative solitons in the complex ginzburg-landau and
  swift-hohenberg equations.
\newblock In {\em Dissipative solitons}, pages 1--17. Springer, 2005.

\bibitem{liehr2013dissipative}
Andreas Liehr.
\newblock {\em Dissipative solitons in reaction diffusion systems}.
\newblock Springer, 2013.

\bibitem{grelu2012dissipative}
Philippe Grelu and Nail Akhmediev.
\newblock Dissipative solitons for mode-locked lasers.
\newblock {\em Nature photonics}, 6(2):84--92, 2012.

\bibitem{grelu2015nonlinear}
Philippe Grelu.
\newblock {\em Nonlinear optical cavity dynamics: from microresonators to fiber
  lasers}.
\newblock John Wiley \& Sons, 2015.

\bibitem{song2019recent}
Yufeng Song, Xujie Shi, Chengfa Wu, Dingyuan Tang, and Han Zhang.
\newblock Recent progress of study on optical solitons in fiber lasers.
\newblock {\em Applied Physics Reviews}, 6(2):021313, 2019.

\bibitem{javaloyes2016dynamics}
J~Javaloyes, P~Camelin, M~Marconi, and M~Giudici.
\newblock Dynamics of localized structures in systems with broken parity
  symmetry.
\newblock {\em Physical review letters}, 116(13):133901, 2016.

\bibitem{ryczkowski2018real}
Piotr Ryczkowski, M~N{\"a}rhi, Cyril Billet, J-M Merolla, Go{\"e}ry Genty, and
  John~Micha{\"e}l Dudley.
\newblock Real-time full-field characterization of transient dissipative
  soliton dynamics in a mode-locked laser.
\newblock {\em Nature Photonics}, 12(4):221--227, 2018.

\bibitem{liu2019revealing}
Xueming Liu and Meng Pang.
\newblock Revealing the buildup dynamics of harmonic mode-locking states in
  ultrafast lasers.
\newblock {\em Laser \& Photonics Reviews}, 13(9):1800333, 2019.

\bibitem{liu2019revealing1}
Xueming Liu, Daniel Popa, and Nail Akhmediev.
\newblock Revealing the transition dynamics from q switching to mode locking in
  a soliton laser.
\newblock {\em Physical review letters}, 123(9):093901, 2019.

\bibitem{peng2019soliton}
Junsong Peng and Heping Zeng.
\newblock Soliton collision induced explosions in a mode-locked fibre laser.
\newblock {\em Communications Physics}, 2(1):1--8, 2019.

\bibitem{ning2014vector}
Qiu-Yi Ning, Hao Liu, Xu-Wu Zheng, Wei Yu, Ai-Ping Luo, Xu-Guang Huang,
  Zhi-Chao Luo, Wen-Cheng Xu, Shan-Hui Xu, and Zhong-Min Yang.
\newblock Vector nature of multi-soliton patterns in a passively mode-locked
  figure-eight fiber laser.
\newblock {\em Optics express}, 22(10):11900--11911, 2014.

\bibitem{wang2018self}
Zhenhong Wang, Zhi Wang, Yange Liu, Ruijing He, Jian Zhao, Guangdou Wang, and
  Guang Yang.
\newblock Self-organized compound pattern and pulsation of dissipative solitons
  in a passively mode-locked fiber laser.
\newblock {\em Optics letters}, 43(3):478--481, 2018.

\bibitem{marconi2014lasing}
Mathias Marconi, Julien Javaloyes, Salvador Balle, and Massimo Giudici.
\newblock How lasing localized structures evolve out of passive mode locking.
\newblock {\em Physical review letters}, 112(22):223901, 2014.

\bibitem{richardson320FsSoliton1991}
D.~J. Richardson, R.~I. Laming, D.~N. Payne, M.~W. Phillips, and V.~J. Matsas.
\newblock 320 fs soliton generation with passively mode-locked erbium fibre
  laser.
\newblock {\em Electronics Letters}, 27(9):730--732, April 1991.

\bibitem{grudininPassiveHarmonicModelocking1993}
A.~B. Grudinin, D.~J. Richardson, and D.~N. Payne.
\newblock Passive harmonic modelocking of a fibre soliton ring laser.
\newblock {\em Electronics Letters}, 29(21):1860--1861, October 1993.

\bibitem{pilipetskii1995acoustic}
AN~Pilipetskii, EA~Golovchenko, and CR~Menyuk.
\newblock Acoustic effect in passively mode-locked fiber ring lasers.
\newblock {\em Optics letters}, 20(8):907--909, 1995.

\bibitem{grudinin1997passive}
AB~Grudinin and S~Gray.
\newblock Passive harmonic mode locking in soliton fiber lasers.
\newblock {\em JOSA B}, 14(1):144--154, 1997.

\bibitem{kutz1998stabilized}
J~Nathan Kutz, BC~Collings, K~Bergman, and WH~Knox.
\newblock Stabilized pulse spacing in soliton lasers due to gain depletion and
  recovery.
\newblock {\em IEEE journal of quantum electronics}, 34(9):1749--1757, 1998.

\bibitem{tangCompoundPulseSolitons2003}
D.~Y. Tang, B.~Zhao, D.~Y. Shen, C.~Lu, W.~S. Man, and H.~Y. Tam.
\newblock Compound pulse solitons in a fiber ring laser.
\newblock {\em Physical Review A}, 68(1):013816, July 2003.

\bibitem{tang2005soliton}
DY~Tang, B~Zhao, LM~Zhao, and Hwa~Yaw Tam.
\newblock Soliton interaction in a fiber ring laser.
\newblock {\em Physical Review E}, 72(1):016616, 2005.

\bibitem{tsatourian2013polarisation}
Veronika Tsatourian, Sergey~V Sergeyev, Chengbo Mou, Alex Rozhin, Vitaly
  Mikhailov, Bryan Rabin, Paul~S Westbrook, and Sergei~K Turitsyn.
\newblock Polarisation dynamics of vector soliton molecules in mode locked
  fibre laser.
\newblock {\em Scientific reports}, 3(1):1--8, 2013.

\bibitem{wang2019optical}
ZQ~Wang, K~Nithyanandan, A~Coillet, P~Tchofo-Dinda, and Ph~Grelu.
\newblock Optical soliton molecular complexes in a passively mode-locked fibre
  laser.
\newblock {\em Nature communications}, 10(1):1--11, 2019.

\bibitem{komarov2009quantization}
Andrey Komarov, Konstantin Komarov, and Fran{\c{c}}ois Sanchez.
\newblock Quantization of binding energy of structural solitons in passive
  mode-locked fiber lasers.
\newblock {\em Physical Review A}, 79(3):033807, 2009.

\bibitem{zhao2009bunch}
LM~Zhao, DY~Tang, Han Zhang, and X~Wu.
\newblock Bunch of restless vector solitons in a fiber laser with sesam.
\newblock {\em Optics express}, 17(10):8103--8108, 2009.

\bibitem{song2016coexistence}
YF~Song, H~Zhang, LM~Zhao, DY~Shen, and DY~Tang.
\newblock Coexistence and interaction of vector and bound vector solitons in a
  dispersion-managed fiber laser mode locked by graphene.
\newblock {\em Optics express}, 24(2):1814--1822, 2016.

\bibitem{wang2018few}
Chong Wang, Lu~Wang, Xiaohui Li, Wenfeng Luo, Tianci Feng, Ying Zhang, Penglai
  Guo, and Yanqi Ge.
\newblock Few-layer bismuthene for femtosecond soliton molecules generation in
  er-doped fiber laser.
\newblock {\em Nanotechnology}, 30(2):025204, 2018.

\bibitem{mou2013bound}
Chengbo Mou, Sergey~V Sergeyev, Aleksey~G Rozhin, and Sergei~K Turitsyn.
\newblock Bound state vector solitons with locked and precessing states of
  polarization.
\newblock {\em Optics express}, 21(22):26868--26875, 2013.

\bibitem{malomed1995bound}
BA~Malomed.
\newblock Bound states in a gas of solitons supported by a randomly fluctuating
  force.
\newblock {\em EPL (Europhysics Letters)}, 30(9):507, 1995.

\bibitem{akhmediev1998stable}
NN~Akhmediev, A~Ankiewicz, and JM~Soto-Crespo.
\newblock Stable soliton pairs in optical transmission lines and fiber lasers.
\newblock {\em JOSA B}, 15(2):515--523, 1998.

\bibitem{grelu2002phase}
Ph~Grelu, F~Belhache, F~Gutty, and J-M Soto-Crespo.
\newblock Phase-locked soliton pairs in a stretched-pulse fiber laser.
\newblock {\em Optics letters}, 27(11):966--968, 2002.

\bibitem{lederer1999multipulse}
MJ~Lederer, Barry Luther-Davies, HH~Tan, C~Jagadish, NN~Akhmediev, and
  JM~Soto-Crespo.
\newblock Multipulse operation of a ti: sapphire laser mode locked by an
  ion-implanted semiconductor saturable-absorber mirror.
\newblock {\em JOSA B}, 16(6):895--904, 1999.

\bibitem{olivier2009origin}
Michel Olivier and Michel Pich{\'e}.
\newblock Origin of the bound states of pulses in the stretched-pulse fiber
  laser.
\newblock {\em Optics express}, 17(2):405--418, 2009.

\bibitem{weill2016noise}
Rafi Weill, Alexander Bekker, Vladimir Smulakovsky, Baruch Fischer, and Omri
  Gat.
\newblock Noise-mediated casimir-like pulse interaction mechanism in lasers.
\newblock {\em Optica}, 3(2):189--192, 2016.

\bibitem{jang2013ultraweak}
Jae~K Jang, Miro Erkintalo, Stuart~G Murdoch, and St{\'e}phane Coen.
\newblock Ultraweak long-range interactions of solitons observed over
  astronomical distances.
\newblock {\em Nature Photonics}, 7(8):657--663, 2013.

\bibitem{he2019formation}
Wenbin He, Meng Pang, Dung-Han Yeh, Jiapeng Huang, CR~Menyuk, and P~St~J
  Russell.
\newblock Formation of optical supramolecular structures in a fibre laser by
  tailoring long-range soliton interactions.
\newblock {\em Nature communications}, 10(1):1--9, 2019.

\bibitem{soto2003quantized}
Jose~M Soto-Crespo, Nail Akhmediev, Ph~Grelu, and F~Belhache.
\newblock Quantized separations of phase-locked soliton pairs in fiber lasers.
\newblock {\em Optics letters}, 28(19):1757--1759, 2003.

\bibitem{zhou2020buildup}
Yi~Zhou, Yu-Xuan Ren, Jiawei Shi, Huade Mao, and Kenneth~KY Wong.
\newblock Buildup and dissociation dynamics of dissipative optical soliton
  molecules.
\newblock {\em Optica}, 7(8):965--972, 2020.

\bibitem{nimmesgern2021soliton}
Luca Nimmesgern, Cornelius Beckh, Hannes Kempf, Alfred Leitenstorfer, and Georg
  Herink.
\newblock Soliton molecules in femtosecond fiber lasers: universal binding
  mechanism and direct electronic control.
\newblock {\em Optica}, 8(10):1334--1339, 2021.

\bibitem{sulimany2018bidirectional}
Kfir Sulimany, Ohad Lib, Gilad Masri, Avi Klein, Moti Fridman, Philippe Grelu,
  Omri Gat, and Hadar Steinberg.
\newblock Bidirectional soliton rain dynamics induced by casimir-like
  interactions in a graphene mode-locked fiber laser.
\newblock {\em Physical review letters}, 121(13):133902, 2018.

\bibitem{casimir1948influence}
Hendrik~BG Casimir and Dirk Polder.
\newblock The influence of retardation on the london-van der waals forces.
\newblock {\em Physical Review}, 73(4):360, 1948.

\bibitem{weill2011spectral}
Rafi Weill, Alexander Bekker, Vladimir Smulakovsky, Baruch Fischer, and Omri
  Gat.
\newblock Spectral sidebands and multipulse formation in passively mode-locked
  lasers.
\newblock {\em Physical Review A}, 83(4):043831, 2011.

\bibitem{bale2009transition}
Brandon~G Bale, Khanh Kieu, J~Nathan Kutz, and Frank Wise.
\newblock Transition dynamics for multi-pulsing in mode-locked lasers.
\newblock {\em Optics express}, 17(25):23137--23146, 2009.

\bibitem{zhao2020observation}
Fengyan Zhao, Hushan Wang, Yishan Wang, Xiaohong Hu, Ting Zhang, and Ran Pan.
\newblock Observation of various bound states based on a graded index multimode
  fiber saturable absorber.
\newblock {\em Laser Physics Letters}, 17(2):025105, 2020.

\bibitem{fu2015passive}
Shijie Fu, Quan Sheng, Xiushan Zhu, Wei Shi, Jianquan Yao, Guannan Shi,
  Robert~A Norwood, and N~Peyghambarian.
\newblock Passive q-switching of an all-fiber laser induced by the kerr effect
  of multimode interference.
\newblock {\em Optics express}, 23(13):17255--17262, 2015.

\bibitem{li2017mode}
Huanhuan Li, Zhaokun Wang, Can Li, Junjie Zhang, and Shiqing Xu.
\newblock Mode-locked tm fiber laser using smf-simf-gimf-smf fiber structure as
  a saturable absorber.
\newblock {\em Optics express}, 25(22):26546--26553, 2017.

\bibitem{wang2017er}
Zhaokun Wang, DN~Wang, Fan Yang, Liujiang Li, Chunliu Zhao, Ben Xu, Shangzhong
  Jin, Shiying Cao, and Zhanjun Fang.
\newblock Er-doped mode-locked fiber laser with a hybrid structure of a
  step-index--graded-index multimode fiber as the saturable absorber.
\newblock {\em Journal of Lightwave Technology}, 35(24):5280--5285, 2017.

\bibitem{wang2018stretched}
Zhaokun Wang, DN~Wang, Fan Yang, Liujiang Li, Chun-Liu Zhao, Ben Xu, Shangzhong
  Jin, Shi-Ying Cao, and Zhan-Jun Fang.
\newblock Stretched graded-index multimode optical fiber as a saturable
  absorber for erbium-doped fiber laser mode locking.
\newblock {\em Optics letters}, 43(9):2078--2081, 2018.

\bibitem{teugin2018all}
U{\u{g}}ur Te{\u{g}}in and B{\"u}lend Orta{\c{c}}.
\newblock All-fiber all-normal-dispersion femtosecond laser with a nonlinear
  multimodal interference-based saturable absorber.
\newblock {\em Optics letters}, 43(7):1611--1614, 2018.

\bibitem{yang2018saturable}
Fan Yang, DN~Wang, Zhaokun Wang, Liujiang Li, Chun-Liu Zhao, Ben Xu, Shangzhong
  Jin, Shi-Ying Cao, and Zhan-Jun Fang.
\newblock Saturable absorber based on a single mode fiber--graded index
  fiber--single mode fiber structure with inner micro-cavity.
\newblock {\em Optics express}, 26(2):927--934, 2018.

\bibitem{chen2018all}
Tao Chen, Qiaoli Zhang, Yaping Zhang, Xin Li, Haikun Zhang, and Wei Xia.
\newblock All-fiber passively mode-locked laser using nonlinear multimode
  interference of step-index multimode fiber.
\newblock {\em Photonics Research}, 6(11):1033--1039, 2018.

\bibitem{zhao2020generation}
Fengyan Zhao, Ning Li, and Hushan Wang.
\newblock Generation of a noise-like pulse from an erbium-doped fiber laser
  based on nonlinear multimode interference.
\newblock {\em Laser Physics}, 30(12):125102, 2020.

\bibitem{zhao2018experimental}
Fengyan Zhao, Hushan Wang, Xiaohong Hu, Yishan Wang, Wei Zhang, Ting Zhang,
  Chuandong Sun, and Zhijun Yan.
\newblock Experimental observation of bound solitons with a nonlinear multimode
  interference-based saturable absorber.
\newblock {\em Laser Physics Letters}, 15(11):115106, 2018.

\bibitem{grudininEnergyQuantisationFigure1992}
A.B. Grudinin, D.J. Richardson, and D.N. Payne.
\newblock Energy quantisation in figure eight fibre laser.
\newblock {\em Electronics Letters}, 28(1):67--68, January 1992.

\bibitem{Tang2005Mechanism}
D.~Y. Tang, L.~M. Zhao, B.~Zhao, and A.~Q. Liu.
\newblock Mechanism of multisoliton formation and soliton energy quantization
  in passively mode-locked fiber lasers.
\newblock {\em Phys. Rev. A}, 72(4):043816, 2005.

\bibitem{weillStatisticalLightmodeDynamics2007}
Rafi Weill, Boris Vodonos, Ariel Gordon, Omri Gat, and Baruch Fischer.
\newblock Statistical light-mode dynamics of multipulse passive mode locking.
\newblock {\em Physical Review E}, 76(3):031112, September 2007.

\bibitem{kelly1992characteristic}
SMJ Kelly.
\newblock Characteristic sideband instability of periodically amplified average
  soliton.
\newblock {\em Electronics Letters}, 28(8):806--807, 1992.

\bibitem{gordonDispersivePerturbationsSolitons1992}
J.~P. Gordon.
\newblock Dispersive perturbations of solitons of the nonlinear
  {{Schr\"odinger}} equation.
\newblock {\em Journal of the Optical Society of America B}, 9(1):91, January
  1992.

\bibitem{haus2000mode}
Herman~A Haus.
\newblock Mode-locking of lasers.
\newblock {\em IEEE Journal of Selected Topics in Quantum Electronics},
  6(6):1173--1185, 2000.

\bibitem{hemanahausModelockingLasers2000}
{Heman A Haus}.
\newblock Mode-locking of lasers.
\newblock {\em IEEE Journal of Selected Topics in Quantum Electronics},
  November/December 2000.

\bibitem{wang2013comparison}
Shaokang Wang, Andrew Docherty, Brian~S Marks, and Curtis~R Menyuk.
\newblock Comparison of numerical methods for modeling laser mode locking with
  saturable gain.
\newblock {\em JOSA B}, 30(11):3064--3074, 2013.

\bibitem{li2010geometrical}
Feng Li, Ping Kong~Alexander Wai, and J~Nathan Kutz.
\newblock Geometrical description of the onset of multi-pulsing in mode-locked
  laser cavities.
\newblock {\em JOSA B}, 27(10):2068--2077, 2010.

\bibitem{soliton2022}
Kfir Sulimany, Offek Tziperman, Yaron Bromberg, and Omri Gat.
\newblock Soliton-pair dynamical transition.
\newblock \url{https://github.com/kfirsuli/Soliton-pair-dynamical-transition},
  2022.

\bibitem{weill2007statistical}
Rafi Weill, Boris Vodonos, Ariel Gordon, Omri Gat, and Baruch Fischer.
\newblock Statistical light-mode dynamics of multipulse passive mode locking.
\newblock {\em Physical Review E}, 76(3):031112, 2007.

\bibitem{finkelstein2021spectral}
Zohar Finkelstein, Kfir Sulimany, Shachar Resisi, and Yaron Bromberg.
\newblock Spectral shaping in a multimode fiber by transmission matrix
  engineering.
\newblock In {\em Frontiers in Optics}, pages JTu1A--116. Optical Society of
  America, 2021.

\bibitem{hausNoiseModelockedLasers1993}
H.A. Haus and A.~Mecozzi.
\newblock Noise of mode-locked lasers.
\newblock {\em IEEE Journal of Quantum Electronics}, 29(3):983--996, March
  1993.

\bibitem{katzNoiseinducedOscillationsFluctuations2010}
Michael Katz, Omri Gat, and Baruch Fischer.
\newblock Noise-induced oscillations in fluctuations of passively mode-locked
  pulses.
\newblock {\em Optics Letters}, 35(3):297--299, February 2010.

\end{thebibliography}
\bibliographystyle{unsrt}

\clearpage
\renewcommand{\theequation}{S\arabic{equation}}
\renewcommand{\thetable}{S\arabic{table}}
\renewcommand{\thefigure}{S\arabic{figure}}

\setcounter{equation}{0}
\setcounter{figure}{0}
\setcounter{table}{0}
\section{Supplementary Information}

\subsection{Experimental setup}

The laser setup is an all-fiber 14m integrated ring cavity, utilizing a 2.7-m Erbium-doped fiber (Thorlabs ER16-8/125) gain medium, based on the design reported in \cite{zhao2020observation}. The EDF is core pumped by a 976-nm diode (Thorlabs BL976-SAG300) laser through a wavelength-division multiplexer (Thorlabs WD9850FA). A polarization-independent optical isolator (Thorlabs IO-H-1550APC) ensures unidirectional lasing, and two fiber polarization controllers (Thorlabs FPC560) tune the overall low cavity birefringence. The nonlinear interference saturable absorber consists of a 1.5m long multimode fiber (OM1, 62.5\,$\mu$m core diameter, $\textrm{NA} =0.275$)) spliced to two single mode fibers (SMF-28) at both ends. A 90/10 fiber coupler (TW1550R2A2) provides the laser output. 

The laser output power is measured by a fast photodiode (Thorlabs DET08CFC) and analyzed with an optical spectrum analyzer (Yokogawa AQ6374) and a real-time $8$GHz oscilloscope (Tektronix MSO70804C).

\subsection{Matlab code for the stochastic numerical simulation}

Here we provide our Matlab code of the physical core utilized to calculate the propagation of the waveform presented in section 2. The complete Matlab package used to generate Figure \ref{transition_sim} is available at \cite{soliton2022}.

The simulation parameters are as follows:
$l=1.636\cdot 10^{-2}$ is the linear loss, $\beta''=-2$ is the group velocity dispersion, $\gamma=4$ is the Kerr coefficient,  $\Omega_g=\sqrt{12}$ is the gain bandwidth, $\delta=0.02$ and $\sigma=0.01$ are the fast saturable absorber constants. $g_d=5\cdot10^{-5}$ is the gain depletion coefficient. $\zeta=10^{-4}$ is the noise amplitude. In Eq. \ref{fluctuating}, $t_R=100000$ is the round trip time. The linear gain is initially $g_0=0.65$ and then $g_0=0.58$ after $8.7\cdot 10^5$ cavity lengths and the saturation power is $P_{sat}=1.8$. We periodically perturb the waveform to account for inhomogeneities and attenuate it by 5\% every cavity length.
\begin{lstlisting}
% Experimental Parameters
PsatTR =1.8; Omega = sqrt(12); g0=0.58; %Gain parameters
Cooupling_ratio = 0.95; CavityLength = 1.45; Loss = 0.2+log(Cooupling_ratio)/CavityLength; % Local and distributed loss
del = 0.02; sigma = 0.01; %Saturaber absorber parameters
gam = 4; beta2 = -2; % Fiber nonlinearity and loss
NOISE_AMP = 0.01; gd = 0.00005; Tr = 100000;

% Numerical Parameters
Nt = 2^12; T = 2^10; dt = T/Nt; 
t = (-Nt/2:1:Nt/2-1)'*dt; dw = 2*pi/T; w = [0:Nt/2-1  -Nt/2:-1]'*dw;
Z = 4000000; h = 0.05; NumSteps = round(Z/h);
SaveDistance = CavityLength*100 ; SaveInterval = round(SaveDistance/h);
raundInterval = round(CavityLength/h);

%Operators
separation = [];
L = (1i*beta2*w.^2-loss)/2;
K = (1 - (w/Omega).^2)/2;

%Initial condition
init_seperations = [40]; 
u0 = 1*exp(-((t-init_seperations(1)/2)/1).^2)...
    +1*exp(-((t+init_seperations(1)/2)/1).^2); 
pulse_index_0 = find(t>-init_seperations(1)/2,1); shift = []; 
averaged_spectrum = u0; Etot = norm(u0).^2*dt; uf = fft(u0); 
uplot = abs(u0).'; uplot_w = abs(fftshift(uf)).'; 
gtplot= [-gd*cumsum(abs(u0).^2)*dt+...
    Etot*gd/Tr*dt.*(1:length(u0))']'; 
Psatf = PsatTR/dt*Nt;
zplot = 0; PULSE_POWER_THREASHOLD = 0.1; WIDTH_THREASHOLD = 50; 

% Propagator
for istep = 1:NumSteps
  uf = uf + sqrt((h/0.05)*(dt/(1/4)))*(1/sqrt(2)*...
      (randn(1,length(uf))'*NOISE_AMP...
      +1i*randn(1,length(uf))'*NOISE_AMP));
  if mod(istep,raundInterval)==0
     uf = uf*Cooupling_ratio;
  end
  g1 = g0/(1+norm(uf)^2/Psatf);   %Saturated gain
  g2 = -2*(g1^2/g0/Psatf)*real(dot(uf,(L + g1*K).*uf));    %frequency dependence
  u=ifft(exp(L*h/2+(g1*h/2+g2/8*h^2)*K).*uf);
  Etot =norm(u).^2*dt;
  Etot_w_noise = Etot;
  [pks,pulse_indexes,width,promince] = ...
      findpeaks(abs(u).^2,'MinPeakProminence'...
      ,PULSE_POWER_THREASHOLD);
  if length(pulse_indexes)>=2
      current_separation = ...
          (abs(pulse_indexes(2)-...
          pulse_indexes(1))*dt);
  end
  if length(pulse_indexes)>=2
      indeces = (1:length(u)); 
      dist_from_pulses = min(abs(indeces - pulse_indexes));
      max_width = max(width);
      noise_avg_pow = ...
          mean(abs(u(dist_from_pulses>...
          (WIDTH_THREASHOLD*max_width))).^2);
      Etot_w_noise = Etot + noise_avg_pow*dt*length(u)*(Tr-T)/T;
      cavity_indexes = ...
          (Tr/2/dt-length(u)/2:...
          Tr/2/dt+length(u)/2-1)';
      gt = -gd*cumsum(abs(u).^2)*dt - ...
          gd*noise_avg_pow*dt*(Tr/2/dt-...
          length(u)/2)+Etot_w_noise*...
          gd/Tr*dt.*cavity_indexes;
  else
      current_separation = 0;
      cavity_indexes = (Tr/2/dt-length(u)/2:...
          Tr/2/dt+length(u)/2-1)';
      gt = -gd*cumsum(abs(u).^2)*dt - ...
          Etot*gd/Tr*dt.*cavity_indexes;
  end
  gt = gt - mean(gt)*T/Tr ;
  uf = fft(exp(+(del+1i*gam)*h*abs(u).^2-...
      sigma*h*abs(u).^4+gt.*h).*u);
  g1 = g0/(1+norm(uf)^2/Psatf);
  uf = exp(L*h/2 + (g1*h/2+g2/8*h^2)*K).*uf;
end
\end{lstlisting}

\subsection{Analytical model for noise mediated interaction mechanism}
The NMI mechanism was first proposed in \cite{weill2016noise}, together with the basic equations of motion governing the pulse dynamics. The interaction is based on fluctuations of pulse timing generated by its overlap with the quasi-CW permeating the cavity. Since NMI is mediated by the gain dynamics, it is expected to act in any multi-pulse mode-locked laser, although its importance in comparison with other interaction mechanisms can vary widely between different systems. We briefly explain the principles of NMI with an emphasis on the aspects relevant to the present work and refer the reader to \cite{weill2016noise} for more details. Notice that here we correct an algebraic mistake in the separation time formula that originally appeared in \cite{weill2016noise}.

\begin{figure}[ht]
\begin{centering}
\includegraphics[width=\columnwidth]{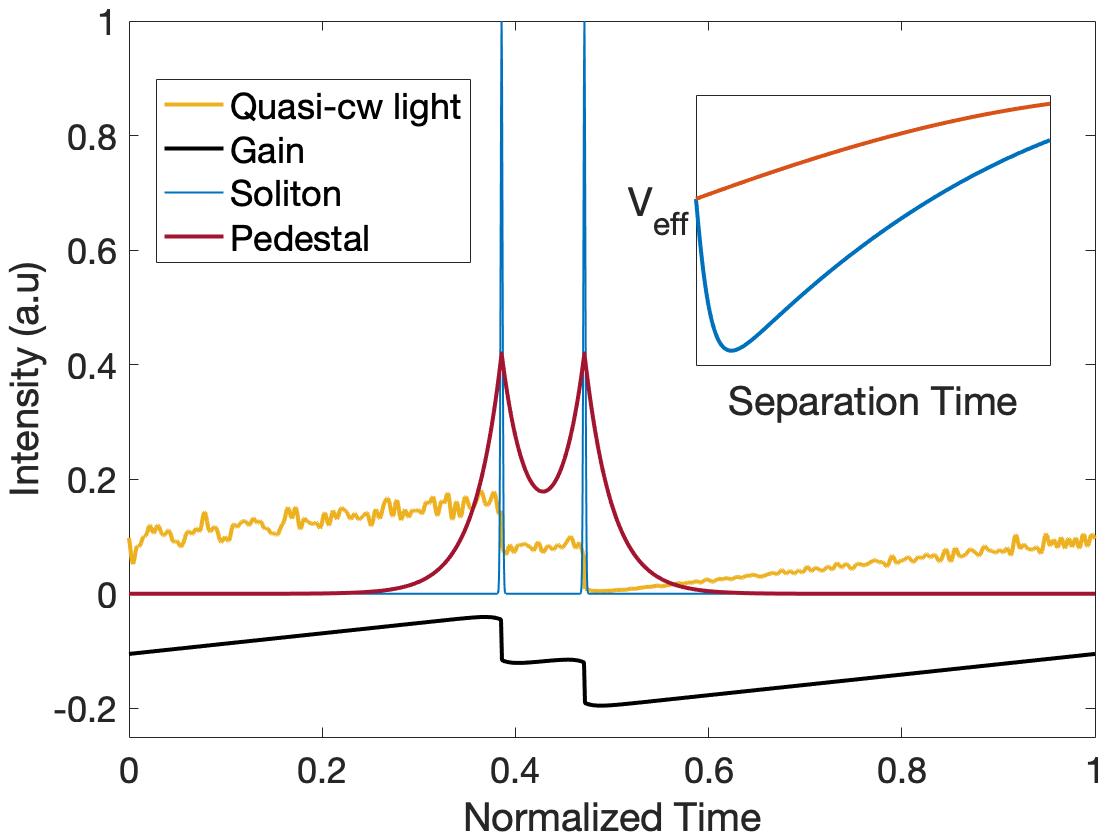}
\par\end{centering}
     \caption{ \textbf{Illustration of the NMI mechanism.} 
     Passage of solitons through the gain medium depletes the instantaneous gain. A pair of pulses yields a smoothed sawtooth gain profile. The quasi-CW light that runs in the cavity is determined by the gain profile, and thus also exhibits a smoothed saw-tooth profile. The NMI is a byproduct of the pulse timing jitter caused by nonlinear overlap interaction between the pulses and the quasi-CW light. The inset shows the effective potential for the noise-mediated interaction of the loosely bound (blue) and tightly bound (red) states.}
 \label{noise_mediated_interaction_explanation}
 \end{figure}

We approximate the laser waveform by an incoherent sum of $n$ non-overlapping soliton waveforms centered at time points $t_k,\;1\le k \le n$ (in the frame moving with the group velocity of a single soliton). Each soliton is accompanied by a weak and broad dispersive-wave pedestal generated by the Gordon-Kelly resonances \cite{weill2011spectral}, and an extended quasi-CW waveform. The quasi-CW is generated by noise buildup, and therefore has a random phase, as indicated in Figure \ref{noise_mediated_interaction_explanation}. The growth of the quasi-CW is limited by the small-signal net loss, and therefore the quasi-CW intensity is inversely proportional to the square root of the net loss in a uniform system. However, since the amplifier gain is depleted by passage of a pulse through the gain medium, and gain recovery is slow, the gain profile $g(t)$ is nonuniform, and therefore the quasi-CW intensity is not uniform either. 

The gain profile is governed by the differential equation
\begin{equation} \label{deplition}
\frac{dg}{dt} = g_r-g_d\sum_k I_p(t-t_k)\ ,
\end{equation}
where $g_r$ and $g_d$ are the gain recovery and gain depletion coefficients, respectively, and $I_p(t)$ is the intensity profile of a soliton with pedestal centered at zero, neglecting the quasi-CW gain depletion, and approximating the gain recovery by a constant as explained following Eq.\ \ref{fluctuating}.

Of the two pulse components, the soliton waveform is approximated by a delta function because its duration is much shorter than all other time scales, and the pedestal waveform is modeled by an exponential decaying on a time scale $w$ that is much longer than the pulse width but much shorter than the cavity round-trip time $t_R$, so that
\begin{equation} \label{pulse}
I_p(t)=E_s\delta(t)+\frac{E_p}{2w}e^{-|t|/w}\ ,
\end{equation}
$E_s$, $E_p$ being the pulse and pedestal energy, respectively. Consequently, the gain profile $g(t)$ is a smoothed-corner sawtooth shaped function, as shown in Figure \ref{noise_mediated_interaction_explanation}, overlaid with the pulse-intensity profile. Note that the gain recovery rate $g_r$, which is fixed in Equation \ref{deplition} by continuity and the total pulse energy, sets the saturation depth of the gain.

The overlap of a pulse and the noisy quasi-CW is a source of timing jitter  \cite{hausNoiseModelockedLasers1993,katzNoiseinducedOscillationsFluctuations2010}, which can be modeled as a diffusion process. The diffusion coefficient $D_k$ of pulse number $k$ is proportional to the instantaneous quasi-CW power at time $t_k$, which is in turn determined by the net gain profile as \cite{weill2016noise},

\begin{equation} \label{difussion}
D_k \propto
(\frac{1}{\sqrt{l-g_k^-}} +
\frac{1}{\sqrt{l-g_k^+}})
\end{equation}
where $l$ is the total small signal loss and $g_k^-$, $g_k^+$ are the values of the gain coefficient just before and just after the $k$th pulse, respectively. Note that the $D_k$ depend on the pulse separations, since the separations determine the gain profile. Consequently, even though the noise itself is symmetrically distributed, the pulse jitter is biased, with an effective noise-induced drift of

\begin{equation} \label{drift}
\frac{d\langle t_k\rangle}{dz} = \frac{1}{2} \frac{\partial D_k}{\partial t_k}\ ,
\end{equation}
where the brackets indicate noise-averaging. 

From this point we consider two-soliton waveforms; the drift equations \ref{drift} can be expressed as overdamped motion in an effective potential \cite{weill2016noise} that is attractive at large separations, but may have a minimum at a separation value $t_s$ of the order of the pedestal width, corresponding to a stable loosely bound state at this separation (see inset of Figure \ref{noise_mediated_interaction_explanation}). 

In a stable steady state configuration, $D_k$ will be maximized with respect to the pulse times $t_k$, as can be seen directly from \ref{drift}. To find the steady state pulse separation we begin by finding the gain profile. We then use \ref{difussion} to find the diffusion coefficients and maximize them with respect to the pulse times. Integrating \ref{deplition}, and finding the recovery rate using the periodic boundary conditions yields
\begin{equation} \label{gain_analytic}
g(t)=g_0 + g_d\int_0 ^t(P-I_p(t'-t_1)-I_p(t'-t_2))dt',
\end{equation}
where $g_0$ is the mean gain and $P$ is the average power of the waveform:
\begin{equation} \label{gain_analytic2}
P=\frac{1}{t_R}\int_0 ^{t_R} (I_p(t'-t_1)+I_p(t'-t_2))dt'
.\end{equation}
Using now equation \ref{pulses} we find the gain profile:
\begin{equation} \begin{gathered} \label{gain_profile}
\frac{l-g(t)}{\bar{g}}=1-\frac{g_d}{\bar{g}}(E_s+E_p) (\frac{2t}{t_R}-1)\\
-\frac{g_d}{\bar{g}}
\begin{cases}
E_s+E_p - \frac{E_p}{2}(e^{\frac{t-t_1}{w}}+e^{\frac{t-t_2}{w}}) & t<t_1\\
\frac{E_p}{2}(e^{\frac{t_1-t}{w}}-e^{\frac{t-t_2}{w}}) & t_1<t<t_2 \\
-(E_s+E_p) + \frac{E_p}{2}(e^{\frac{t_1-t}{w}}+e^{\frac{t_2-t}{w}}) & t_2<t
\end{cases}
\end{gathered}
\end{equation}
where $\bar{g}$ is absolute value of the mean net gain. We can assume without loss of generality that the pulse times obey $t_R=t_1+t_2$, and define $t_s=t_2-t_1$. The relation between the gain and the diffusion coefficients \ref{difussion} gives the following expression
\begin{equation}
\begin{gathered} \label{Diffusion_coef}
D_1\propto \frac{1}{\sqrt{1+\frac{g_d}{\bar{g}}(\frac{t_s}{t_R}(E_s+E_p)+\frac{E_p}{2}(-1+e^{-\frac{t_s}{w}}))}}\\
+\frac{1}{\sqrt{1+\frac{g_d}{\bar{g}}((\frac{t_s}{t_R}-1)(E_s+E_p)+\frac{E_p}{2}(1+e^{-\frac{t_s}{w}}))}}\\
D_2\propto \frac{1}{\sqrt{1+\frac{g_d}{\bar{g}}(-\frac{t_s}{t_R}(E_s+E_p)+\frac{E_p}{2}(1-e^{-\frac{t_s}{w}}))}}\\
+\frac{1}{\sqrt{1+\frac{g_d}{\bar{g}}((1-\frac{t_s}{t_R})(E_s+E_p)-\frac{E_p}{2}(1+e^{-\frac{t_s}{w}}))}}.\\
\end{gathered}
\end{equation}
The diffusion coefficients in the model do not depend on the absolute pulse timings but only on the separation $t_s$, thus \ref{drift} in a steady state turns into:
\begin{equation}
    \frac{\partial{D_s}}{\partial{t_s}}=0,\quad \textrm{where $D_s=D_1+D_2$},\
\end{equation}
from here, some algebra leads to
\begin{equation}
    \tilde{t}_s=\tilde{w} \ln(\frac{E_p t_R}{2(E_p+E_s)w})=\tilde{w} \ln(\frac{\tilde{E}_p}{2\tilde{w}}),\
\end{equation}
where we introduced the notations $\tilde{E}_p=\frac{E_p}{E_p+E_s}$, $\tilde{t}_s=\frac{t_s}{t_R}$ and $\tilde{w}=\frac{w}{t_R}$ which completes the derivation of equation \ref{separation}.
\end{document}